\documentclass[a4paper,nodate]{sab} 

\usepackage{graphicx} 
\usepackage{xcolor}
\usepackage{txfonts,natbib} 
\bibpunct{(}{)}{;}{a}{}{,} 
\usepackage{hyperref}
\usepackage{url}
\usepackage[T1]{fontenc} 
\usepackage{t1enc} 
\renewcommand\ackname{Acknowledgements.} 
\renewcommand\andname{\&}
\renewcommand\keywordname{Keywords.} 
\renewcommand\figureptname{Figure}
\renewcommand\tableptname{Table}

\renewcommand\noteaddname{Note added to proofs.}
\renewcommand\andname{\&}

\idline{Boletim da Sociedade Astron\^omica Brasileira, {\bf 31}, no. 1, XX-XX}{3}

\def\msun{M$_\odot$\/}

\def\mbh{$M_{\rm BH}$\/}

\begin{document} 
\title{What drives the Coronal Lines?} 
\subtitle{Resolving the forbidden, high-ionization emission regions in a sample of AGNs}

\authorrunning{S. Panda et al.}\titlerunning{What drives the Coronal Lines?}

\author{S. Panda \inst{1}, A. Rodr\'iguez-Ardila \inst{1,2}, M. A. Fonseca-Faria \inst{1}, F. C. Cerqueira-Campos \inst{2}, M. Marinello \inst{1} \and L. G. Dahmer-Hahn \inst{3}} 
\institute{Laborat\'orio Nacional de Astrof\'isica - Rua dos Estados Unidos 154, Bairro das Na\c c\~oes. CEP 37504-364, Itajub\'a, MG, Brazil\\ \email{spanda@lna.br} \and Divis\~ao de Astrof\'isica, Instituto Nacional de Pesquisas Espaciais, Avenida dos Astronautas 1758, S\~ao Jos\'e dos Campos, 12227-010, SP, Brazil  \and Shanghai Astronomical Observatory, Chinese Academy of Sciences, 80 Nandan road, Shanghai 200030, China}
\date{Received } 

\Abstract {Emission-line studies in the active galactic nuclei (AGNs), particularly those utilizing high spatial resolution, provide the most accurate method to determine critical quantities of the central engine and of the gas a few tens of parsecs away. Using seeing-limited data with spectroscopy, we have explored the extended narrow-line region for a sample of active galactic nuclei (AGNs) with strong, forbidden emission lines that have high-ionization potentials (IP $\gtrsim$ 100 eV). We have studied the optical and near-infrared spectra for these AGNs, extracted and compared their spectral energy distributions, and put constraints on the physical conditions of the region producing the coronal lines. We have realized a novel black hole mass scaling relation with one such prominent coronal line - [Si {\sc vi}] 1.963$\mu$m, over the 10$^6$ - 10$^8$ \msun{} interval, that suggests photoionization by the continuum produced by the accretion disk is the primary physical process at play here. We perform a detailed parameter space study to optimize the emission from these coronal lines in terms of fundamental black hole parameters and test predictions that can be used to measure the kinematics of the extended X-ray emission gas. With the successful launch and first light of the JWST, we are well-poised to refine our findings using the superb angular resolution of the telescope that will allow us to map the inner few parsecs to the central supermassive black holes. This opens up the study of the higher ionization lines that will be spatially resolved by JWST, expanding our sample to tens of hundreds of AGNs, and putting firmer constraints on the physical conditions in the coronal line region.}{Estudos de linha de emissão nos núcleos ativos de galáxias (AGNs), particularmente aqueles que utilizam alta resolução espacial, fornecem o método mais preciso para determinar quantidades críticas do motor central e do gás a algumas dezenas de parsecs de distância. Usando dados de espectroscopia limitada por $seeing$, exploramos a região estendida de linhas estreitas em uma amostra de AGNs com fortes linhas de emissão proibidas que têm alto potencial de  ionização (IP $\gtrsim$ 100 eV). Estudamos espectros no óptico e no infravermelho próximo para esses AGNs, extraímos e comparamos suas distribuições espectrais de energia e colocamos restrições nas condições físicas da região que produz as linhas coronais. Percebemos uma nova relação de escala para a massa de buraco negro com uma dessas linhas coronais proeminentes - [Si {\sc vi}] 1.963$\mu$m, no intervalo de 10$^6$ - 10$^8$ \msun{}. Esse resultado sugere a fotoionização pelo contínuo produzido pelo disco de acreção como o processo físico primário dominante. Realizamos um estudo detalhado do espaço de parâmetros que otimizam a emissão dessas linhas coronais em termos de propriedades fundamentais de buracos negros e previsões de modelos que podem ser usadas para medir a cinemática do gás de emissão de raios-X estendido. Com o lançamento bem-sucedido e a primeira luz do JWST, estamos bem posicionados para refinar nossas descobertas usando a excelente resolução angular do telescópio. Isso nos permitirá mapear os parsecs mais internos dos buracos negros supermassivos centrais. Nossos resultados abrem caminho para o estudo das linhas de mais alta ionização que serão resolvidas espacialmente pelo JWST, expandindo nossa amostra para dezenas de centenas de AGNs e colocando vínculos mais firmes nas condições físicas na região da linhas coronais.}

\keywords{(Galaxies:) quasars: emission lines -- Methods: observational -- Techniques: spectroscopic -- Methods: data analysis -- Accretion, accretion disks -- Radiation mechanisms: thermal}


\maketitle 

\section{A brief introduction to coronal emission lines in AGNs}

Coronal lines (CLs) spread over the X-rays, optical and IR spectrum. Although often fainter than the classical medium-ionization lines used for photoionization diagnosis, high angular resolution in nearby AGNs has shown that CLs particularly in the near-IR are among the most conspicuous features in active galaxies \citep[][]{2006ApJ...653.1098R, 2011ApJ...739...69M, 2017MNRAS.470.2845R, 2021MNRAS.500.2666C}.

\section{A novel black hole mass scaling using coronal lines in Type-1 active galaxies}

\begin{figure*}
    \centering
    \includegraphics[width=0.75\textwidth]{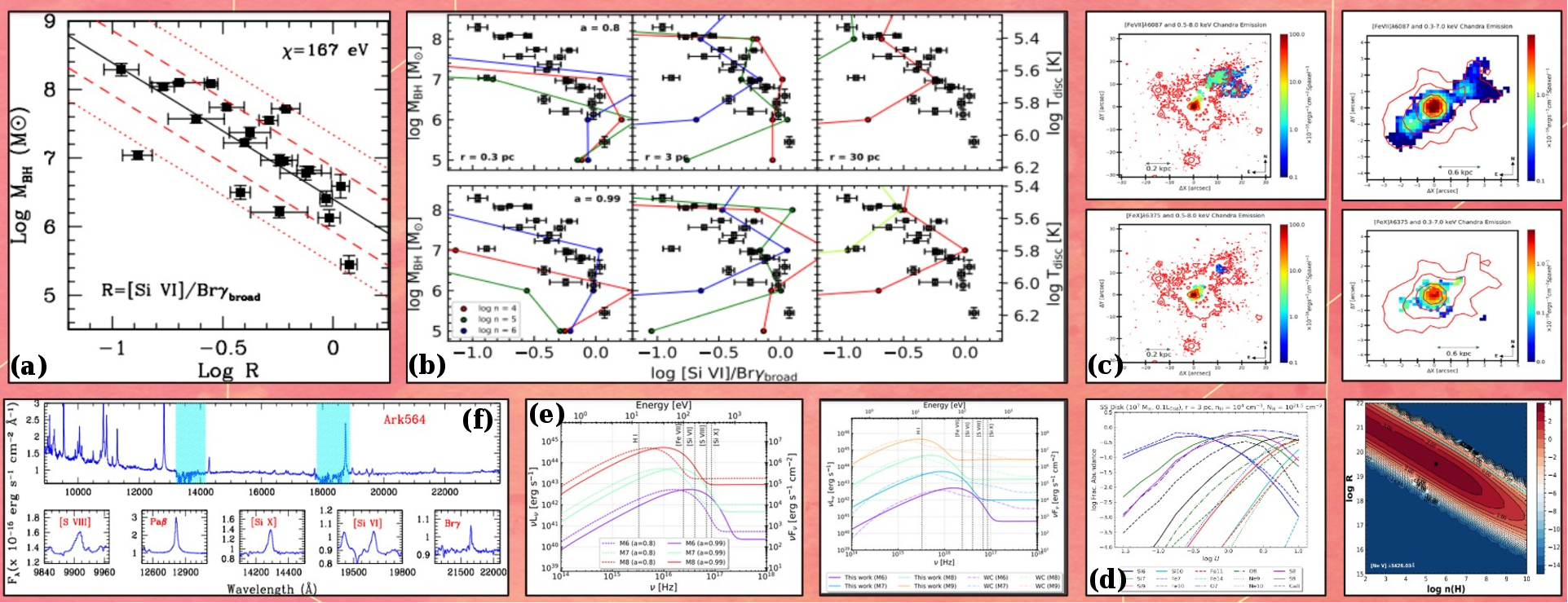}
    \caption{Collage of the various published and soon-to-be published works from our group focused on the study and analysis of coronal emission in active galaxies. (clockwise from the upper left panel:) (a) observed [Si {\sc vi}]/Br$\gamma_{\rm broad}$ vs. \mbh{} correlation; (b) CLOUDY predictions for [Si {\sc vi}]/Br$\gamma_{\rm broad}$ vs \mbh{} using a standard disk ionising continuum for a range of physical parameters. Observed data points are shown using black squares; (c) Emission maps of the Circinus (left panels) and IC 5063 (right panels), [\ion{Fe}{vii}]~$\lambda$6087~\AA\ (top) and [\ion{Fe}{x}]~$\lambda$6375~\AA\ (bottom); (d, left) CLOUDY photoionization model predictions for the fractional abundances of relevant ionization states of iron and silicon and H- and He-like oxygen and neon. (right) Density histogram for [\ion{Fe}{vii}]~$\lambda$6087~\AA, referenced to the incident continuum at 5100\AA; (e) Generic AGN ionizing continua (with and without warm comptonization). All curves are normalised to 10\% of the Eddington limit; and (f) NIR spectra of Ark~564 (obtained using GNIRS). The shaded areas mark regions of bad atmospheric transmission. }
    \label{fig:collage}
\end{figure*}

Objects in our work presented in \cite{2022MNRAS.510.1010P} are selected on the basis of having black hole (BH) masses determined by reverberation mapping and single epoch optical and/or near-IR spectra with accurate CL measurements. The CLs used in the analysis are - [Fe\,{\sc vii}]~$\lambda$6087\AA\ in the optical and [S\,{\sc viii}]~0.991$\mu$m, [Si\,{\sc x}]~1.432$\mu$m and [Si\,{\sc vi}]~1.964$\mu$m in the near-IR. They are among the strongest CLs in AGNs \citep{2003MNRAS.343..192R, 2011ApJ...743..100R, 2017MNRAS.467..540L} and span a wide IP range, 100-350 eV. In addition,  H\,{\sc i} lines of H$\beta$, Pa$\beta$ and Br$\gamma$ are employed. The whole set samples the ionizing continuum over the 13.6-351 eV range. An exemplary NIR spectrum from our sample is shown in Figure \ref{fig:collage}f for a high accreting, low mass BH (Ark~564, $\sim$10$^{6.59}$\;M$_{\odot}$). Using bonafide BH mass estimates and the line ratio [Si\,{\sc vi}]/Br$\gamma_{\rm broad}$ as a genuine tracer of the AGN ionizing continuum, a BH-mass scaling relation for a BH mass range 10$^6$ - 10$^8$ M$_{\odot}$, is found for a sample of 21 reverberation-mapped AGNs (see Figure \ref{fig:collage}a). The dependence follows a linear regression in log-scale: \textbf{log ${\rm M_{BH}}$ = (6.40$\pm$0.17) - (1.99$\pm$0.37)$\times$log$\left(\rm{\frac{[Si\; VI]}{Br\gamma_{\rm broad}}}\right)$}, with a dispersion in BH mass of 0.47 dex that is already contesting the classical M$_{\rm BH}$ - $\sigma_{\star}$ relation (the latter has a scatter $\sim$0.44 dex). With the assumption of a geometrically thin, optically thick accretion disk as the dominant component of the ionizing continuum (Figure \ref{fig:collage}e), and a suitable range of local cloud densities, $n_{\rm e} \sim 10^{4-6}$ cm$^{-3}$ and cloud distances, 0.3 pc $\lesssim r \lesssim$ 30 pc, for the survival of CLs, photoionization models using {\sc CLOUDY}, \citep{2017RMxAA..53..385F} provide a fair representation of the $M_{\rm BH}$ and [Si {\sc vi}]/Br$\gamma_{\rm broad}$ correlation (Figure \ref{fig:collage}b).

\section{Extended coronal emission as a kinematical tracer of the X-ray gas}

We report the first characterization of an extended outflow of high ionized gas in the Circinus Galaxy and in IC 5063 (Figure \ref{fig:collage}c) by means of the CLs [\ion{Fe}{vii}]~$\lambda$6087~\AA\ and [\ion{Fe}{x}]~$\lambda$6375~\AA. In both cases, the [\ion{Fe}{vii}] emission is located within the ionization cone already detected in the [\ion{O}{iii}]~$\lambda$5007~\AA\ line and is found to extend up to a distance of $\sim$700 pc (Circinus) and $\sim$1.2 kpc (IC 5063) from the centre. Correspondingly, the [\ion{Fe}{x}]~$\lambda$6375~\AA\ emission extends up to $\sim$400 pc (Circinus) and $\sim$700 pc (IC 5063). The physical conditions of the gas show that the extended coronal emission is likely the remnant of shells inflated by the passage of a radio jet. This scenario is supported by extended X-ray emission, which is spatially coincident with the morphology and extension of the [\ion{Fe}{vii}] and [\ion{Fe}{x}] gas \citep[see][for more details]{2020ApJ...895L...9R}.

\section{Photoionization modelling of the coronal line emission - predictions for JWST}

From the theoretical viewpoint, we use photoionization models to predict the physical conditions of high-ionization footprint emission lines in the optical and near-IR that can be used to accurately trace the kinematics and physical conditions of active galactic nucleus-ionized, X-ray emission-line gas \citep[following the prescription of][]{2022MNRAS.511.1420T}. The footprint lines are formed in gas over the same range in ionization state as the H- and He-like of O and Ne ions, which are also the source of X-ray emission lines. The footprint lines can be detected with optical and IR telescopes, such as the Hubble Space Telescope/STIS and JWST/NIRSpec, and can potentially be used to measure the kinematics of the extended X-ray emission gas (Figure \ref{fig:collage}d-left).

\section{Onto the future}

A direct extension of these published works is to realize a comprehensive database of coronal lines (see e.g., Figure \ref{fig:collage}d-right) in a diverse population of active galaxies which in turn would aid in the detection of newer AGNs, especially at the higher redshift regime with the upcoming JWST. An important aspect of the usage of these coronal lines' diagnostics would be in the search and discovery of the accreting black holes that belong to the intermediate mass range, i.e. the Intermediate Mass Black Holes (IMBHs). With the successful deployment of the JWST, we are well poised to test these predictions for the coronal line emission and their imminent connection with the central accreting BH.

\begin{acknowledgements} S.P. and A.R.A. acknowledge financial support from the CNPq grants 164753/2020-6 and 300936/2023-0, and 312036/2019-1, respectively.  \end{acknowledgements} 

\bibliographystyle{mnras}
\bibliography{references}

\end{document}